**Title:** Timescales in Time-to-Event Analyses in Pregnancy

**Authors:** Chase D. Latour,[1] Shalini S. Ramachandra,[1] Daniel A. Xu[1], Elizabeth Diemer[1,2] Viktor H. Ahlqvist[3,4], Kristin Palmsten[5], Stefanie Hinkle[1,6], Sunni L. Mumford[1,6], Enrique Schisterman[1,6], Jessie K. Edwards[7,8], Ellen C. Caniglia[1,6]

**Affiliations:**
[1] Department of Biostatistics, Epidemiology, and Informatics, University of Pennsylvania, Philadelphia, PA, USA
[2] Department of Pediatrics, Children's Hospital of Philadelphia, Philadelphia, PA, USA
[3] Department of Medical Epidemiology and Biostatistics, Karolinska Institutet, Stockholm, Sweden
[4] Department of Biomedicine, Aarhus University, Aarhus, Denmark
[5] Pregnancy and Child Health Research Center, HealthPartners Institute, Minneapolis, Minnesota, USA
[6] Department of Obstetrics and Gynecology, University of Pennsylvania, Philadelphia, PA, USA
[7] Department of Epidemiology, University of North Carolina at Chapel Hill, Chapel Hill, NC, USA
[8] Carolina Population Center, University of North Carolina at Chapel Hill, Chapel Hill, NC, USA

**Conflict of Interest:** CDL has received payment from Headwater Science (previously NoviSci, which was acquired by Target RWE), Regeneron Pharmaceuticals, and Amgen for work unrelated to this project. KP has research contracts with AbbVie, Biogen, GSK, Pfizer, and Sanofi. VHA reports speaker fees to Angelini Pharma and the Organization of Teratology Information Specialists outside the submitted work. EFS reports speaker and consulting fees from Merck not related to this work.

**Funding:** PCORI Contract # ME-2023C2-3334.

**Abstract**

The target trial framework is increasingly used to define the causal estimand for an observational analysis. However, designing time-to-event analyses that align with the causal estimand is challenging in pregnancy because the timescale on which outcomes are often defined, “gestational age”, differs from the timescale on which outcomes are measured, “time since study entry”. We argue that neither timescale is “best” but that each aligns with a different estimand. In this article, we use the target trial framework to define the causal estimands corresponding to each timescale. We consider a (hypothetical) target trial comparing randomization to 17-alpha hydroxyprogesterone caproate (17-OHPC) versus placebo between 16-20 weeks of gestation to prevent delivery before 37 weeks’ gestation, whether by miscarriage, stillbirth, or live birth. We define the estimand of time-to-event analyses on the “time since study entry” timescale as the effect of randomization to 17-OHPC versus placebo between 16-20 weeks of gestation on delivery before 37 weeks’. In contrast, the causal estimand of time-to-event analyses using the “gestational age” timescale corresponds to the effect had, counter to fact, all participants been randomized at one gestational age (e.g., 16 weeks). We review specific challenges to estimation and possible solutions for each timescale.

## Introduction

The target trial framework emphasizes the importance of defining a study's causal estimand, or the causal question it aims to answer.[1–14] Applying this framework illustrates what quantities a given dataset can be used to estimate and the necessary assumptions to do so. However, designing analyses that align with a target trial can be challenging in perinatal studies where the timescale on which the outcome is defined—"gestational age"—may not be aligned with "time since study entry," a common analytic timescale. These challenges are compounded when using time-to-event statistical methods to address right censoring.

In time-to-event analysis, the timing of outcomes is captured in relation to time zero, the point at which eligibility criteria are applied, treatment is assigned, and follow-up begins.[15,16] "Time since study entry" corresponds to the timescale wherein time zero is study entry, or the timepoint at which follow-up begins for a person (e.g., randomization), and the time-to-event is the time between study entry and the outcome. This timescale is well suited to settings where investigators are interested in the timing of events after study entry.[17] However, in perinatal studies, investigators are often interested in whether an outcome occurred within a specific gestational age range, not necessarily whether it occurred within a specified time after study entry. For example, preterm birth is defined by the gestational age at delivery (e.g., <37 weeks'),[18] not the time elapsed after treatment start. Therefore, "gestational age" may offer a more interpretable timescale than time since study entry for many pregnancy-related outcomes.

Shifting time-to-event analyses to the "gestational age" timescale, wherein time zero is some observed gestational age and the time-to-event is the time elapsed since that age, avoids the misalignment associated with the "time since study entry" timescale. It also matches the fetuses-at-risk approach, which argues for using gestational age as survival time, thereby including all living fetuses at that gestational age in the denominator for risk.[19–21] There are numerous articles detailing how to implement such analyses in pregnancy,[22–24] but there has been little attention to their corresponding causal estimands.

Neither the "gestational age" nor "time since study entry" timescales are inherently "best"; instead, each targets a different causal question.[25] This article uses the target trial framework to explicitly define these causal estimands for a hypothetical study in which we are interested in estimating the effect of randomization to treatment versus placebo on the risk of delivery before 37 weeks' gestation. Motivated by this example, we clarify, first, that the estimand of time-to-event analyses using the "time since study entry" timescale corresponds to the effect of randomization within an observed gestational age range (corresponding to the gestational age range at which individuals enrolled in the study). We also discuss how dichotomizing pregnancy outcomes by gestational age (e.g., <37 vs ≥37 weeks) presents analytic challenges to estimating this effect. Second, we describe how the estimand shifts when we use "gestational age" as the analytic timescale, instead estimating the effect had everyone been randomized at that specific gestational age (e.g., the earliest gestational age at which individuals enrolled in the study). This approach aligns the end of follow-up across participants but introduces variable left truncation.

## Motivating Example

We motivate this discussion with a hypothetical trial modeled on prior work.[26,27] Imagine we enrolled people who were ≥18 years of age with a singleton pregnancy between 16 weeks and 0 days of gestation ($16^{0/7}$ weeks) and $20^{6/7}$ weeks' gestation and history of a previous spontaneous preterm birth. Participants were randomized to receive weekly intramuscular injections of 17-alpha hydroxyprogesterone caproate (17-OHPC) or placebo from enrollment through $36^{6/7}$ weeks' gestation or delivery, whichever occurred first, and the primary outcome was a composite outcome of delivery before $37^{0/7}$ weeks' gestation, whether by miscarriage, stillbirth, or live birth. Participants were followed from randomization until end of pregnancy, 37 weeks', or loss to follow-up (LTFU), whichever occurred first. For simplicity, we only consider the intention-to-treat estimand and focus on estimating risks at study end (i.e., after all pregnancies have ended or reached 37 weeks').

## "Time Since Study Entry" as a Timescale

The "time since study entry" timescale aligns with the hypothetical trial described above. Implementing a time-to-event analysis comparing individuals randomized to 17-OHPC versus placebo using this timescale would involve treating study entry (at $16^{0/7}$-$20^{6/7}$ weeks' gestation) as time zero and defining persons' times-to-event as the time between entry and the composite outcome, 37 weeks' gestation, or LTFU, as shown in **Figure 1A**.[15,16] The intention-to-treat estimate would address the following question: what is the effect of being randomized to initiate 17-OHPC (versus placebo) between $16^{0/7}$ and $20^{6/7}$ weeks' gestation on the risk of fetal death or live birth before 37 weeks'? The answer would inform decisions for the population of pregnant people with a history of spontaneous preterm birth and a distribution of gestational ages (and other effect measure modifiers) at entry into prenatal care that mirrors that of the trial.

### *Causal Estimands and Unintuitive Competing Events*

**Table 1** defines the causal estimand for this analysis: the *total* effect of randomization to 17-OHPC versus placebo between $16^{0/7}$ and $20^{6/7}$ weeks' gestation on the composite outcome, through all causal paths.[28–30] Those paths include the effect of randomization to 17-OHPC versus placebo on lengthening gestation beyond 37 weeks and anything else that treatment causes, such as hospitalization for an adverse drug event, that also causes or prevents delivery before 37 weeks'.

Estimating the total effect would be straightforward if no participants were right censored (e.g., LTFU). We would simply estimate risk as the proportion of participants who experienced the outcome among those randomized to 17-OHPC and placebo, respectively, and contrast these risks (**Figure S1**). In a scenario with right censoring, however, such an approach may be invalid. For example, fitting a logistic regression for the binary outcome among those who remain uncensored is not expected to return an unbiased estimate without additional analytic approaches or strong assumptions about censoring (**Discussion S1**). Instead, we can apply time-to-event estimators to estimate these treatment effects if, counter to fact, no participants were right censored, thereby estimating the estimand from our target trial.

Analytic challenges arise, however, when conducting time-to-event analyses on the "time since study entry" timescale: some people reach 37 weeks' gestation before the official end of follow-up (i.e., before *everyone* has delivered or reached 37 weeks'), and how we handle these participants determines what we estimate. Censoring people at 37 weeks' gestation implies that they could go on to have the event at some later, unobserved time.[15,31] If all participants entered the study at the same gestational age, censoring at term would not result in a biased estimate of the total effect. However, censoring at term biases estimates when participants enter a study at different gestational ages.[32–36] Consider the hypothetical data in **Figure 1A**: person 1 enrolled at 16 weeks' gestation and experienced a preterm birth at 34 weeks, while person 2 enrolled at 20 weeks and delivered at 38 weeks (a term birth). Censoring person 2 upon reaching term, 17 weeks after study entry, would lead a time-to-event estimator to impute a risk of the composite outcome for that person at 18 weeks after study entry, although they were no longer being at-risk (**Figure 2A**). Such an analysis targets the controlled direct effect of randomization to 17-OHPC versus placebo on the risk of the composite outcome under an intervention that prevents participants from reaching term.[28,29,37,38] One could argue that this is an uninformative estimate since we do not want to prevent reaching term, or 37 weeks' gestation.

A more appropriate alternative would be to treat reaching term as a competing event, or an event that precludes delivery before 37 weeks' gestation.[37,38] Competing events are common in pregnancy studies,[28,36,39,40] the most intuitive being pregnancy loss.[21,28,36,39–45] Term delivery is a less intuitive one, but for pregnancy outcomes defined by dichotomizing gestational age at delivery, such as our composite outcome, use of the "time since study entry" timescale can lead to reaching term functioning as a competing event.

To instead estimate the total effect of 17-OHPC on our composite outcome, we have two options. The first is to use methods that do not treat the competing event of "reaching term" as a censoring event, as demonstrated below (**Figure 2B**).[37,38] The second is to align the at-risk periods such that we can end follow-up for all participants at 37 weeks' gestation. One such approach would involve implementing the time-to-event analyses stratified by gestational age at randomization and then estimating the risk in the baseline population as an information-weighted average of those stratified risks.[11,14] Such an approach may reduce precision in adjusted analyses, as it allows the relationship between all covariates and the treatment or outcome to vary by gestational age, and is not immune to other competing events such as "healthy" live birth (**Discussion S2**).[36,40] Another approach to align these at-risk periods is to conduct analyses on the "gestational age" timescale; however, as discussed later, this targets a different causal estimand.

### ***Applied Illustration***

We provide a simple, worked example to illustrate the importance of "reaching term" (37 weeks' gestation) as a competing event in our example study. Analyses were performed with R statistical software version 4.5.2 (R Core Team, Vienna, Austria); the code and results are included in **Supplemental File 1** (with comments and documentation written by Claude Code [4.7 Opus, Anthropic, San Francisco, CA, USA] under supervision by CDL).

We generated counterfactual outcomes under placebo and 17-OHPC for 1,000 simulated observations, under a scenario in which 17-OHPC increased the length of gestation by 2 weeks (**Figure 3, Table S1)**. We introduced non-informative censoring due to LTFU at 18.5 weeks from study entry such that the participants remaining uncensored had the same risks as censored persons, resulting in 10% LTFU. We derived the true risks and risk difference (RD) for the total effect from the uncensored counterfactual outcome data: 400/1,000 participants (40%) experienced the composite outcome under 17-OHPC versus 600/1,000 (60%) under placebo, for a RD of -20%. We then compared the true RD to RD estimates derived from analyses that do and do not treat reaching term as a censoring event.

We first estimated risks with the Kaplan-Meier estimator,[37,38,46] with time-to-event defined as the number of days between study entry and the composite outcome, LTFU, or 37 weeks' gestation. Because 37 weeks gestation occurs at different days since study entry for persons who entered at different gestational ages, this approach corresponds to treating both LTFU and reaching term as censoring events. Under this construction, we estimated the risk as 47% under 17-OHPC and 60% under placebo, providing a RD of -13%, different from the true RD of -20%.

In the second analysis, we estimated risks in the right-censored counterfactual outcome data with the Aalen-Johansen estimator,[47] an extension of the Kaplan-Meier estimator that non-parametrically estimates risk in the presence of multiple outcome types (i.e., "reaching term" and the composite outcome).[37,38] This approach does not treat reaching term as a censoring event, effectively setting the hazard of the composite outcome to zero once a participant reaches term. When using this estimator, we estimated the same risks and RD as the truth (17-OHPC: 40%, placebo: 60%, RD: -20%). This simple case study illustrates that censoring pregnancies at term in time-to-event analyses may result in biased estimates of the total effect.

**"Gestational Age" as a Timescale**

To avoid the misalignment between the "time since study entry" timescale and the timescale on which outcomes are often defined, investigators may analyze the hypothetical trial's data on the "gestational age" timescale, wherein we treat a specific gestational age as time zero, and the timing of outcomes reflects the time since that gestational age. Such an analysis is illustrated in **Figure 1B**, where we use the first observed gestational age at trial entry ($16^{0/7}$ weeks' gestation) as time zero for all participants, and end follow-up for all participants at 37 weeks'. Here, we can see that the timing of study entry and exit reflect the gestational ages at trial entry and at the study outcome or 37 weeks' gestation. Further, the end of the at-risk period is the same for all participants (i.e., 37 weeks'), thereby avoiding bias due to censoring at term.

Using "gestational age" as the analytic timescale estimates a different causal estimand than our hypothetical trial, as shown in the right-side column of **Table 1**. Explicitly, such an analysis aims to answer the question: what is the effect of randomization to 17-OHPC versus placebo at $X$ weeks' gestation, where $X$ can take on any observed value but is common for all participants?[48] Put differently, such an analysis seeks to estimate the effect if, counter to fact, all participants who factually enrolled in our trial had actually entered at a specific gestational age, for example, $16^{0/7}$ weeks'.

**Figure 1B** illustrates a key challenge with this type of analysis: gestational weeks 16 through 19 are observed for person 1 but not person 2. In fact, person 2 would not have been eligible for this hypothetical trial based on their factual gestational age at trial entry. Including person 2, then, would require setting time zero as 16 weeks' gestation but then looking ahead in time (i.e., to 20 weeks) to determine person 2's actual treatment assignment, which induces immortal time.[49–55] In addition, person 2 could only enter the study if they remained pregnant through that risk period, inducing selection bias. Correspondingly, we could say that this sample is subject to variable left truncation with regard to our target trial (**Discussion S3**): the start of follow-up varies across participants according to their factual gestational age at study entry.[48]

While multiple estimators have been proposed to address time-varying selection due to survival such as the extended Kaplan-Meier and Aalen-Johansen estimators,[23,55,56] (discrete-time) Cox model,[48,57–59] and *g*-computation,[60–63] these models, as typically implemented, assume that late entry is random or non-informative, such that observed risks among participants who entered early are representative of those who entered later, and that the treatment effect is not modified by gestational age at randomization.[23,48,57,64,65] When left truncation is informative (i.e., late entry is related to other variables such as healthcare engagement) or a modifier of the treatment effect estimate, investigators must layer additional approaches on top of these estimators to remove the bias (**Discussion S4**).[55,56,60,61,66–71]

### *Applied Illustration*

We include a simple, worked example to illustrate the causal estimand targeted by analyses on the "gestational age" timescale (**Supplemental File 2**). Code for this analysis was generated with Claude Code (4.7 Opus, Anthropic, San Francisco, CA, USA) under supervision by one of the authors (CDL).

Again, we simulated data for 1,000 hypothetical participants. **Figure 4** shows counterfactual outcome data for 10 pregnant persons who *could* have enrolled in our trial if they had been identified at 16 weeks' gestation. Each observation in the figure represents 100 simulated participants (**Tables S2**). **Figures 4A-4B** present their counterfactual outcomes had all 1,000 (10 observations in the figure) been randomized to 17-OHPC or to placebo at 16 weeks' gestation. **Figures 4C-4D** illustrate participants' counterfactual outcomes had all been randomized to 17-OHPC or placebo at their observed (factual) gestational ages at trial entry. We created these counterfactual outcomes first by specifying outcomes and times-to-event under randomization to placebo at 16 weeks' gestation. We then created the treated counterfactuals by assuming that 17-OHPC lengthened gestation by 2 weeks for all deliveries previously occurring after 20 weeks' gestation, regardless of gestational age at initiation. Five hundred participants (five observations in the figure) factually entered the trial at 16 weeks' gestation, while 400 (four observations) factually entered at 20 weeks'. The remaining 100 (1 observation) were excluded because they experienced miscarriage at 19 weeks' gestation, precluding trial entry at 20 weeks'. Left truncation was non-informative: those who entered the study late had the same outcome risk as those who entered at 16 weeks.

Based on the counterfactual outcomes, the true risks had all potential participants entered the trial at 16 weeks' gestation were 200/1,000 (20%) under 17-OHPC versus 600/1,000 (60%) under placebo, corresponding to a RD of -40%. In contrast, the true risks based on factual gestational age at trial entry were 100/900 (11%) under 17-OHPC and 500/900 (56%) under placebo (RD: -45%). We can see here that these true risks and RDs differ because one of the potential participants did not factually enter the trial.

We next conducted the analysis on the "gestational age" timescale. We considered only those counterfactual outcomes after factual trial entry (i.e., **Figures 4C-4D**). We used the extended Kaplan-Meier estimator to account for participants who entered "late" (i.e., at 20 weeks' gestation). For this analysis, we created a variable that indicated the gestational age at the start of follow-up (i.e., 16 or 20 weeks' gestation) and the gestational age at the end of follow-up (i.e., the outcome, LTFU, or 37 weeks' gestation). Further, we created a binary outcome indicator as described above. Using this model, we estimated risk of the composite outcome as 20% under 17-OHPC versus 60% under placebo (RD: -40%), which aligns with the true risks and RD from the analysis where all participants enrolled at 16 weeks' gestation but *not* the analysis using the counterfactual outcomes after factual trial entry.

**Left Truncation and Competing Events**

To this point, we have discussed the challenges of each analytic timescale separately. However, researchers may sometimes aim to answer questions that require accounting for both left truncation and competing events. Grappling with these issues jointly will require that investigators carefully consider their target causal estimand.[28,29] To illustrate, consider that we are interested in quantifying the total effect of being randomized to 17-OHPC versus placebo at 16 weeks' gestation on the risk of preterm *live* birth, that is, the effect through all causal paths, including randomization's effect on both miscarriage and stillbirth before 37 weeks' gestation. In this case, we would want to implement a time-to-event estimator that treated miscarriage as a competing event while simultaneously accounting for late entry, which could be accommodated, for example, with the "extended" Aalen-Johansen estimator under strong assumptions (in particular, that left truncation is non-informative conditional on measured covariates).[23]

**Discussion**

Misalignment between the timescale on which the outcome is defined and the timescale of data collection is pervasive within perinatal research, in large part because outcomes of interest are often defined by dichotomizing gestational age (e.g. preterm birth is defined as gestational age at delivery <37 weeks).[23,48,57,64,72] We describe pregnancy-specific issues of misalignment between these timescales within the target trial framework and formalize the causal estimands that correspond to estimates from time-to-event analyses on each timescale.[8,10,73,74] This discussion is particularly timely given the increasing use of time-to-event estimators in pregnancy for study designs such as clone-censor-weighting[75–77] and sequential nested trial emulation[11,14,78] that rely on artificial right censoring to avoid immortal time bias.[7,9,11,14,79]

The data presented in this article are simple, toy examples that were contrived to illustrate (1) that "reaching term" can function as a competing event on the "time since study entry" timescale and (2) that using the "gestational age" timescale targets a different causal estimand than a target trial that allows randomization at different gestational ages. However, the extent to which these analytic choices result in differing estimates will depend on the data. To point (1), for example, the degree of bias induced by censoring pregnancies at term will depend on the frequency of the competing event. If no pregnancies reach term, then there will be no bias. Otherwise, risk estimates will be biased upward and may be less precise than those from analyses that do not censor at term.[40] To point (2), for example, bias due to left truncation will depend on how much the target population and left truncated populations differ in the distribution of effect measure modifiers, as well as the probability of the outcome or pregnancy loss between the targeted gestational age and the time of study entry. While bias may be minimal in some cases, constructing time-to-event analyses to align with the causal estimand will avoid these issues.

## Conclusions

This article leverages the target trial framework to clarify the causal estimands of perinatal time-to-event analyses on the "time since study entry" versus "gestational age" timescale. We demonstrated that neither timescale is "best" but that each targets a different causal question. For our (hypothetical) target trial, using the "time since study entry" timescale targets the effect of randomization within a specified gestational age window at enrollment, while using the "gestational age" timescale targets the effect if, counter to fact, all participants had been randomized at a specific gestational age. When constructing analyses for pregnancy studies with right-censored data, investigators must carefully consider which estimand they intend to target and the analytic challenges specific to that estimand. If the outcome is defined by occurrence before a specific gestational age, analyses on the "time since study entry" timescale will lead reaching that specified gestational age to function as a competing event. In contrast, analyses on the "gestational age" timescale avoid that concern but require approaches that account for variable left truncation.

## TABLES AND FIGURES

**Figure 1.** Illustration of the (A) "time from study entry" and (B) "gestational age" timescales in our hypothetical trial of 17-alpha hydroxyprogesterone caproate (17-OHPC) versus placebo for risk of pregnancy loss, stillbirth, or live birth <37 weeks' gestation. The end of a person's at-risk period (37 weeks') is represented by a vertical dashed line. The blue box in (B) shows person-time that is unobserved for person 2 but observed for person 1 (i.e., "immortal"). For each timescale, we include example analytic data for time zero (i.e., the start of follow-up), the outcome (i.e., fetal death or live birth < or ≥37 weeks'), and the time-to-event (i.e., weeks from time zero to the outcome or 37 weeks' gestation, whichever occurs first).

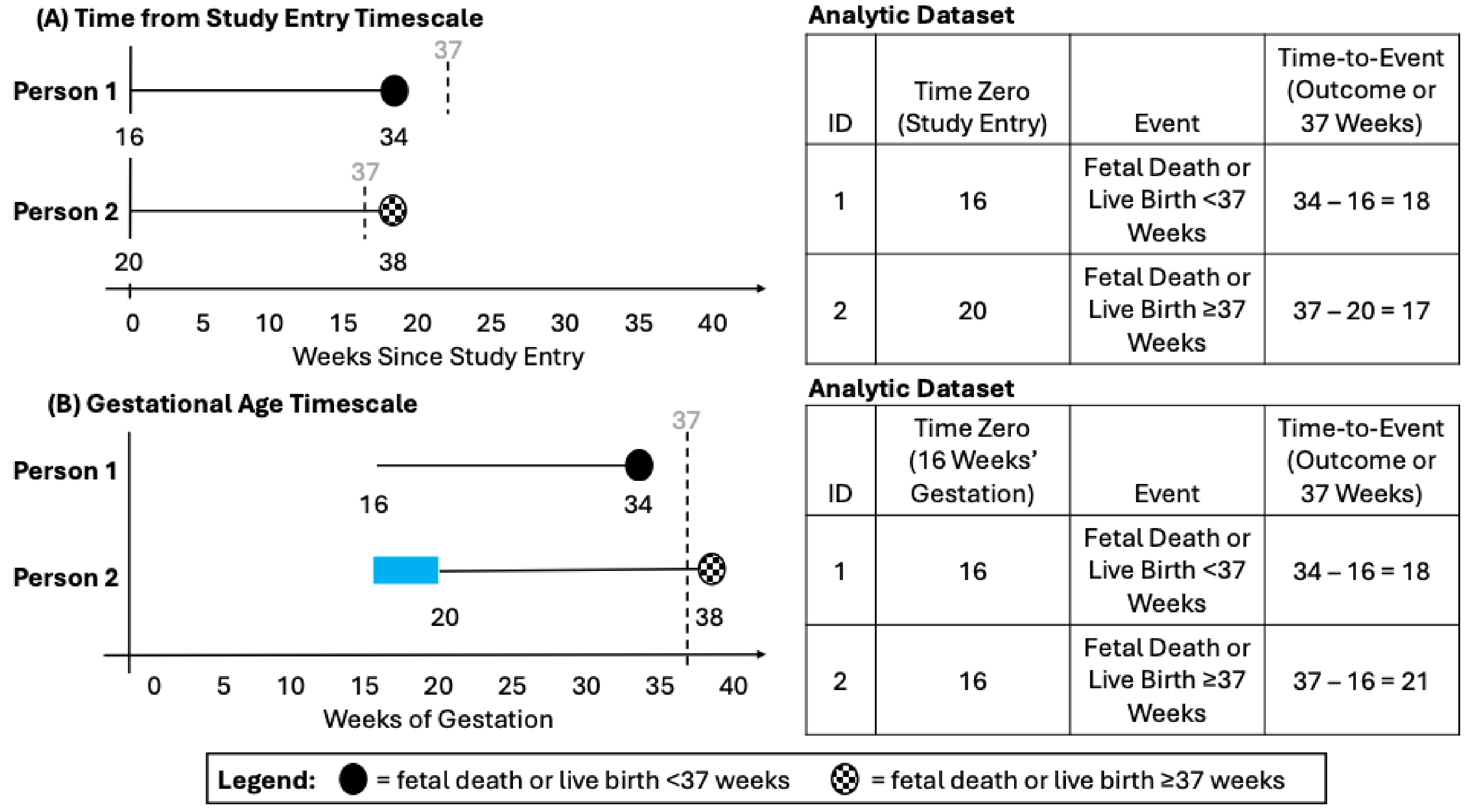

**Analytic Dataset**

| ID | Time Zero (Study Entry) | Event | Time-to-Event (Outcome or 37 Weeks) |
|---|---|---|---|
| 1 | 16 | Fetal Death or Live Birth <37 Weeks | 34 – 16 = 18 |
| 2 | 20 | Fetal Death or Live Birth ≥37 Weeks | 37 – 20 = 17 |

**Analytic Dataset**

| ID | Time Zero (16 Weeks' Gestation) | Event | Time-to-Event (Outcome or 37 Weeks) |
|---|---|---|---|
| 1 | 16 | Fetal Death or Live Birth <37 Weeks | 34 – 16 = 18 |
| 2 | 16 | Fetal Death or Live Birth ≥37 Weeks | 37 – 16 = 21 |

**Table 1.** Key protocol components for two target trials in which we analyze data from a hypothetical trial comparing 17-α-hydroxyprogesterone caproate (17-OHPC) versus placebo for risk of preterm birth on the "time from study entry" and "gestational age" analytic timescales in time-to-event analyses. The target trial implied by the "time from study entry" trial aligns with the hypothetical trial in our motivating example.

| Protocol Component | Target Trial when using "Time from Study Entry" as the Analytic Timescale | Target Trial when using "Gestational Age" as the Analytic Timescale |
|---|---|---|
| **Eligibility Criteria** | Inclusion Criteria:<br>• Viable, singleton pregnancy between $16^{0/7}$ and $20^{6/7}$ weeks of gestation<br>• ≥18 years of age<br>• Spontaneous preterm birth in previous pregnancy | Inclusion Criteria:<br>• Viable, singleton pregnancy<br>• X weeks of gestation, where X can take on any value but must be common for all participants (e.g., 16 weeks' gestation)<br>• ≥18 years of age<br>• Spontaneous preterm birth in previous pregnancy |
| **Treatment Strategies** | Treatment: Initiate and continue once-weekly injections of 17-OHPC through $36^{6/7}$ weeks' gestation or delivery, whichever occurs first.<br><br>Comparator: Initiate and continue once-weekly injections of placebo through $36^{6/7}$ weeks' gestation or delivery, whichever occurs first. | Same. |
| **Assignment Procedures** | Random assignment in a 2:1 ratio. | Same. |
| **Follow-up Period** | Patients are followed from randomization until the first of delivery or loss to follow-up. | Same. |
| **Outcome** | Fetal death or live birth before 37 weeks of gestation. | Same. |
| **Causal Contrasts of Interest**[a,b] | Intention-to-treat effect: The effect of initiating 17-OHPC versus placebo between $16^{0/7}$ and $20^{6/7}$ weeks of gestation.<br><br>$\Pr(Y_{21}^{a=1} - Y_{21}^{a=0})$ | Intention-to-treat effect: The effect of initiating 17-OHPC versus placebo at X weeks of gestation.<br><br>$\Pr(Y_{37-X}^{a=1,g=X} - Y_{37-X}^{a=0,g=X})$ |

[a] Consider that we are interested in estimating the treatment effect as a risk difference. Let $A$ denote treatment arm (1: 17-OHCP, 0: placebo), $G$ the gestational age at trial entry (16, 17, 18, 19, or 20), and $Y_k$ the study outcome before week $k$ (1: fetal death or live birth before 37 weeks of gestation, 0: any other outcome). For each person, $Y_k^a|G = g$ indicates the counterfactual outcome by week $k$ under randomization to treatment $A = a$ conditional upon entering the trial at the observed gestational age, and $Y_k^{a,g}$ indicates the potential outcome by week $k$ under randomization to treatment $A = a$ at $G = g$ weeks of gestation. X is the same as described in "Eligibility Criteria".

[b] The estimate of $\Pr(Y_{21}^{a=1} - Y_{21}^{a=0})$ would technically be derived from the following summation: $= \sum_g \Pr(Y_{21}^{a=1} - Y_{21}^{a=0}|G = g) \Pr(G = g)$. This reflects the fact that the estimate derived from the trial would reflect the estimated risk difference among a population of pregnant people with the same gestational age at the treatment decision as our trial population at trial entry..

**Figure 2.** Data from Figure 1A stopping follow-up at term (37 weeks' gestation). Surviving the risk period is treated as (A) a censoring event and (B) a competing event. Individual 1 experienced a live or stillbirth at 34 weeks' gestation, while individual 2 carried their pregnancy through 37 weeks' gestation.

**(A) Censoring pregnancies at 37 weeks' gestation.**

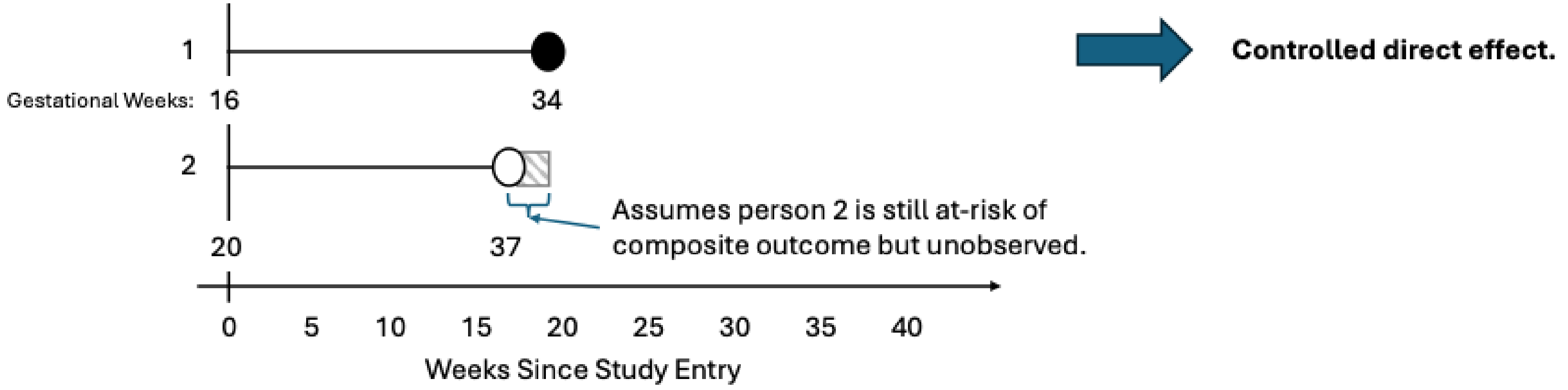

**(B) Reaching 37 weeks' gestation as a competing event.**

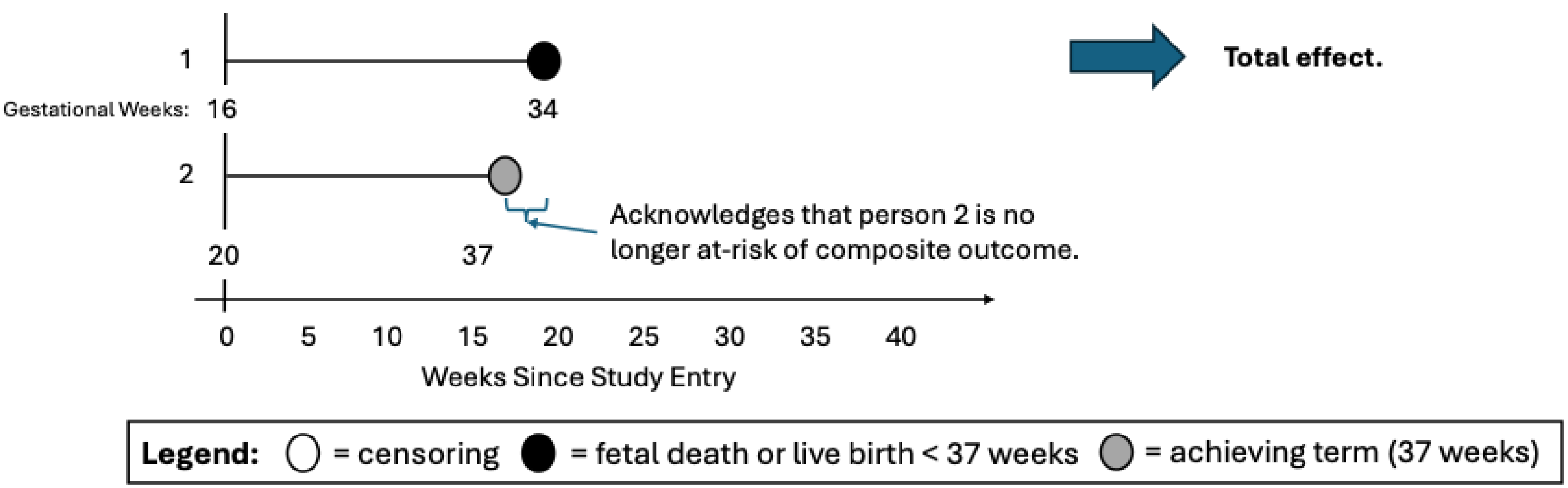

**Figure 3.** Counterfactual outcome data for 10 participant profiles in our hypothetical trial under randomization to (A) placebo versus (B) 17-alpha hydroxyprogesterone caproate (17-OHPC) between 16 and 20 weeks of gestation. Each profile represents 100 participants in the simulated data. Participant profile identifier (ID) and gestational age (GA) at entry are indicated to the left of each figure. Occurrence of the composite outcome (delivery before 37 weeks' gestation, whether by miscarriage, stillbirth, or live birth) is indicated with a circle, while deliveries at ≥37 weeks' gestation are indicated with a star. Observed outcomes are represented with solid, black shapes, while unobserved outcomes are indicated with grey shapes. Loss to follow-up is indicated by a transition from solid, black to a dashed, grey horizontal line. Numbers under the outcomes indicate the GA at the outcome. Vertical grey dotted lines represent 37 weeks' gestation.

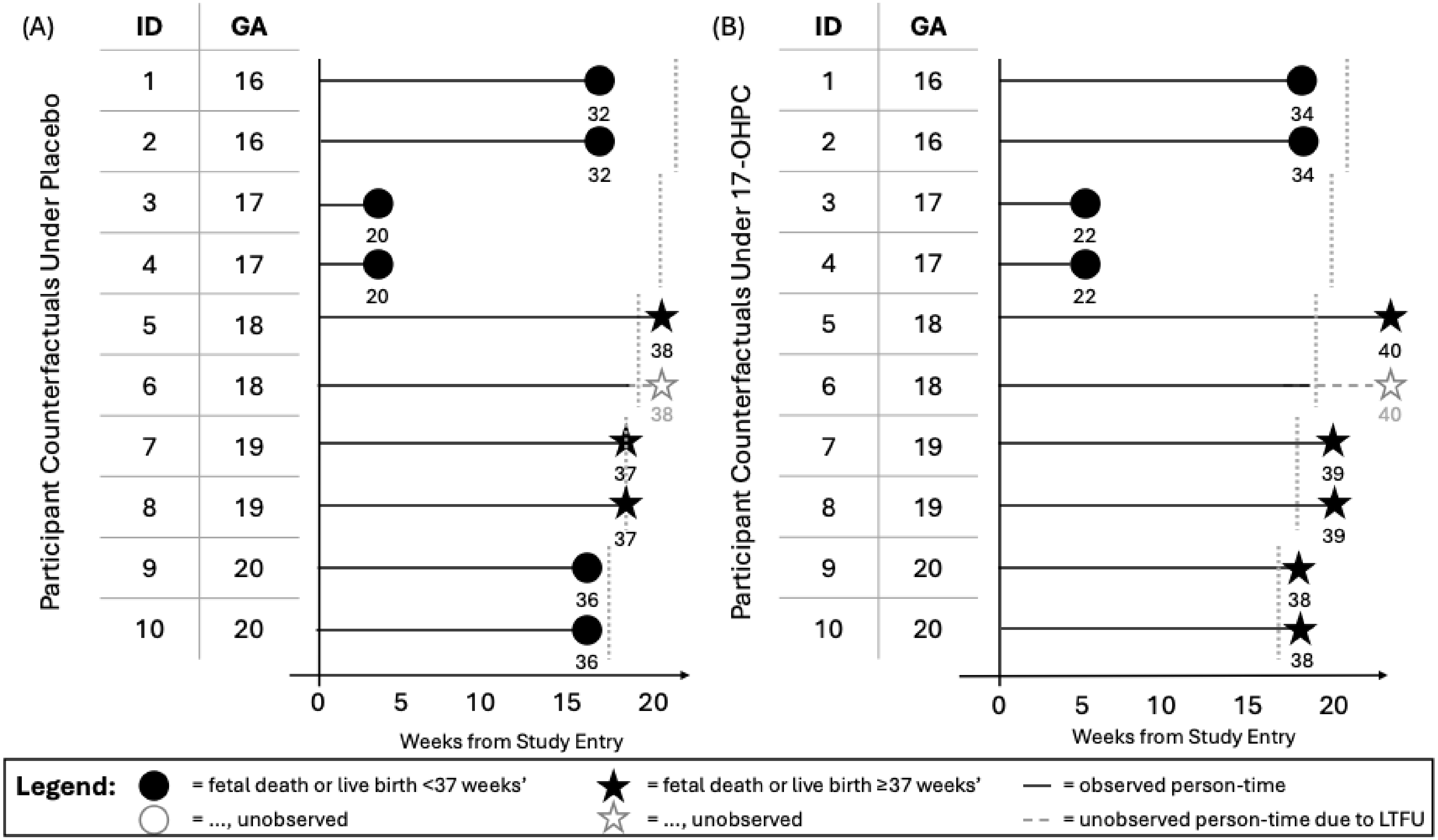

**Figure 4.** Counterfactual outcome data for 10 participant profiles in our hypothetical example trial under randomization to (A) placebo at 16 weeks' gestation, (B) 17-alpha hydroxyprogesterone caproate (17-OHPC) at 16 weeks of gestation, (C) placebo at factual trial entry, and (D) 17-OHPC at factual trial entry. Each profile represents 100 participants in the simulated data. Participant profile identifier (ID) and gestational age (GA) at entry are indicated to the left of each figure. Occurrence of the composite outcome is indicated with a circle, while deliveries at ≥37 weeks' gestation are indicated with a star. Numbers on the x-axis and under the lines and outcome shapes indicate gestational age at the event. Vertical grey dotted lines represent 37 weeks' gestation.

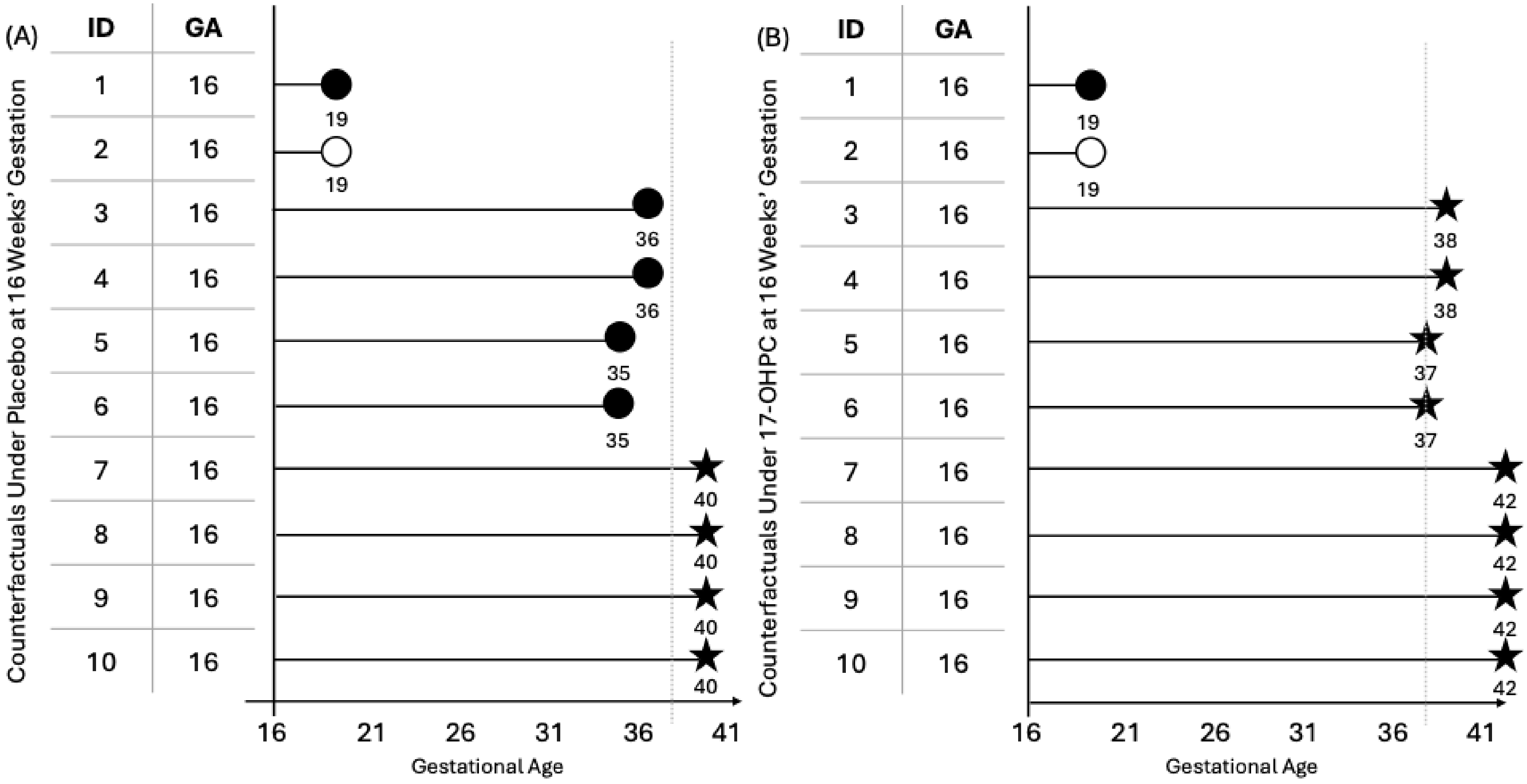

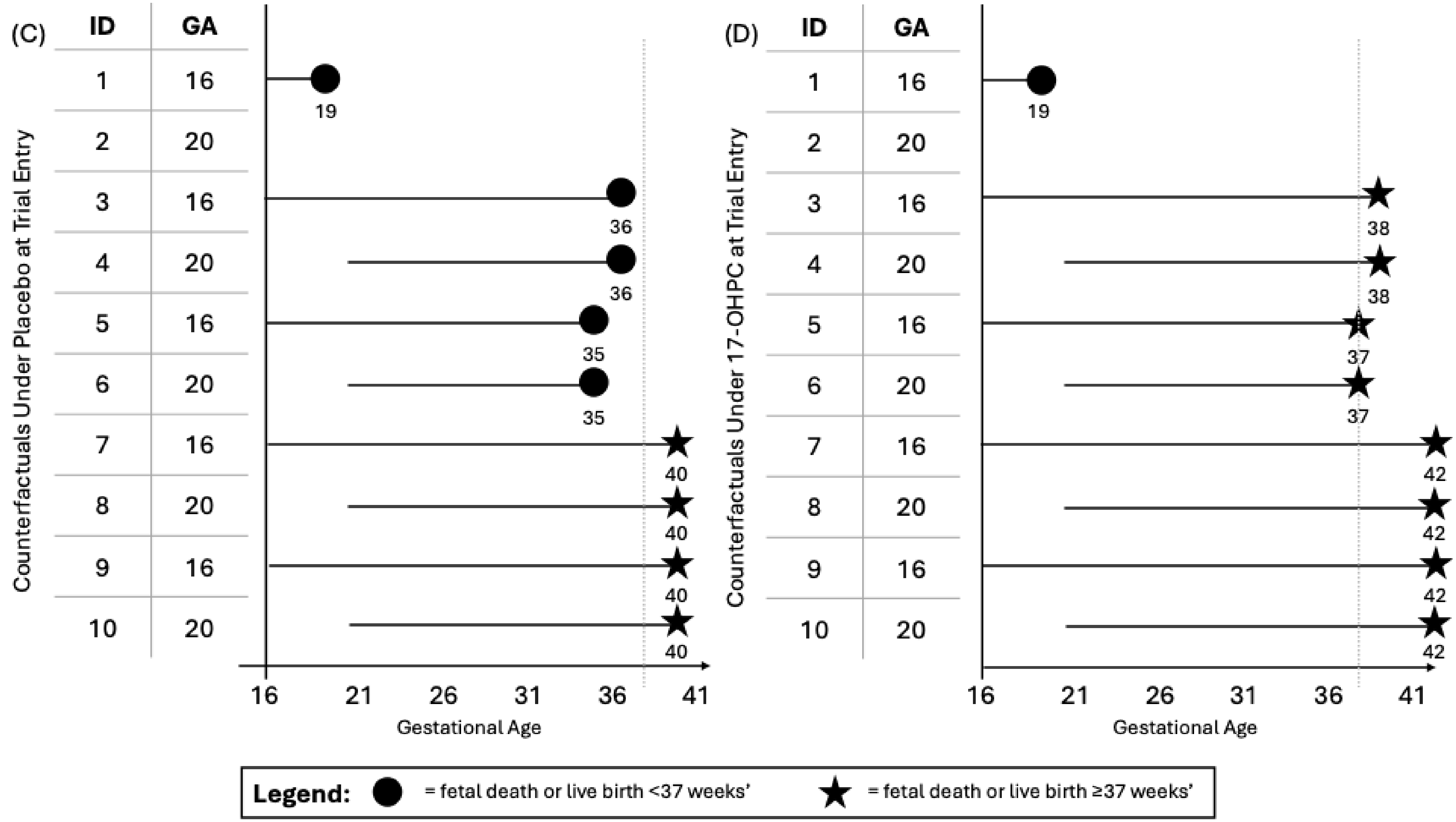

(C)
ID
GA
1
16
2
20
3
16
4
20
5
16
6
20
7
16
8
20
9
16
10
20
19
36
36
35
35
40
40
40
40
Counterfactuals Under Placebo at Trial Entry
16
21
26
31
36
41
Gestational Age
(D)
ID
GA
1
16
2
20
3
16
4
20
5
16
6
20
7
16
8
20
9
16
10
20
19
38
38
37
37
42
42
42
42
Counterfactuals Under 17-OHPC at Trial Entry
16
21
26
31
36
41
Gestational Age
Legend:
= fetal death or live birth <37 weeks'
= fetal death or live birth ≥37 weeks'

**SUPPLEMENT**

## Table of Contents

## Figures

**Figure S1.** Estimating the risk of miscarriage, stillbirth, or live birth (<37 weeks' gestation) among treated and untreated participants in a trial where no participants are lost to follow-up.

**Outcomes Among Treated**

| ID | Gestational Age at Study Entry | Gestational Age at Miscarriage, Stillbirth, or Live Birth | Miscarriage, Stillbirth, or Live Birth < 37 Weeks' |
|---|---|---|---|
| 1 | 16 | 22 | 1 |
| 2 | 18 | 24 | 1 |
| 3 | 19 | 28 | 1 |
| 4 | 19 | 32 | 1 |
| 5 | 16 | 36 | 0 |
| 6 | 17 | 37 | 0 |
| 7 | 17 | 38 | 0 |
| 8 | 20 | 38 | 0 |
| 9 | 18 | 40 | 0 |
| 10 | 20 | 40 | 0 |

**Overall Risk of Composite Outcome (Treated):**
4/10 = 0.4

**Outcomes Among Untreated**

| ID | Gestational Age at Study Entry | Gestational Age at Miscarriage, Stillbirth, or Live Birth | Miscarriage, Stillbirth, or Live Birth < 37 Weeks' |
|---|---|---|---|
| 1 | 16 | 22 | 1 |
| 2 | 18 | 24 | 1 |
| 3 | 19 | 28 | 1 |
| 4 | 19 | 32 | 1 |
| 5 | 16 | 36 | 1 |
| 6 | 17 | 37 | 1 |
| 7 | 17 | 38 | 0 |
| 8 | 20 | 38 | 0 |
| 9 | 18 | 40 | 0 |
| 10 | 20 | 40 | 0 |

**Overall Risk of Composite Outcome (Untreated):**
6/10 = 0.6

**Risk Difference:** 0.4 – 0.6 = -0.2
**Risk Ratio:** 0.4/0.6 = 0.67

## Tables

**Table S1.** Analytic data corresponding to the 10 participant profiles in Figure 3, including the true outcome data, the binary event indicator and time-to-event when censoring at 37 weeks' gestation, and the multinomial outcome indicator and time-to-event when achieving term (37 weeks' gestation) is treated as a competing event.

| | Placebo | | | | | | 17-OHPC | | | | | |
|---|---|---|---|---|---|---|---|---|---|---|---|---|
| | True Outcome Status | | Censoring At 37 Weeks' | | 37 Weeks as Competing Event | | True Outcome Status | | Censoring At 37 Weeks' | | 37 Weeks as Competing Event | |
| **ID** | **Event[a]** | **Time to Event** | **Event[a]** | **Time to Event** | **Event[a]** | **Time to Event** | **Event[a]** | **Time to Event** | **Event[a]** | **Time to Event** | **Event[a]** | **Time to Event** |
| **1** | 1 | 16 | 1 | 16 | 1 | 16 | 1 | 18 | 1 | 18 | 1 | 18 |
| **2** | 1 | 16 | 1 | 16 | 1 | 16 | 1 | 18 | 1 | 18 | 1 | 18 |
| **3** | 1 | 3 | 1 | 3 | 1 | 3 | 1 | 5 | 1 | 5 | 1 | 5 |
| **4** | 1 | 3 | 1 | 3 | 1 | 3 | 1 | 5 | 1 | 5 | 1 | 5 |
| **5** | 0 | 20 | 0 | 19 | 2 | 19 | 0 | 22 | 0 | 19 | 2 | 19 |
| **6** | 0 | 20 | 0 | 18.5 | 0 | 18.5 | 0 | 22 | 0 | 18.5 | 0 | 18.5 |
| **7** | 0 | 18 | 0 | 18 | 2 | 18 | 0 | 20 | 0 | 18 | 2 | 18 |
| **8** | 0 | 18 | 0 | 18 | 2 | 18 | 0 | 20 | 0 | 18 | 2 | 18 |
| **9** | 1 | 16 | 1 | 16 | 1 | 16 | 0 | 18 | 0 | 17 | 2 | 17 |
| **10** | 1 | 16 | 1 | 16 | 1 | 16 | 0 | 18 | 0 | 17 | 2 | 17 |

[a] Event indicator values are 0, corresponding to composite outcome not observed; 1, composite outcome observed; and 2, reached term, or 37 weeks' gestation.

**Table S2.** Analytic data corresponding to the 10 participant profiles in Figure 4, including their counterfactual outcomes if (4A) all had been randomized to placebo at 16 weeks' gestation, (4B) all had been randomized to 17-OHPC at 16 weeks' gestation, (4C) all had been randomized to placebo at factual gestational age at trial entry, and (4D) all had been randomized to 17-OHPC at their factual gestational age at trial entry.

| | | **True Counterfactual Outcomes Under…** | | | | **Counterfactual Outcomes Included in Analytic Dataset for Extended Kaplan-Meier Estimator** | | | | | |
|---|---|---|---|---|---|---|---|---|---|---|---|
| | | **Randomization @ 16 Weeks to …** | | **Randomization at Factual GA at Trial Entry to…** | | **Under Randomization to Placebo** | | | **Under Randomization to 17-OHPC** | | |
| **Participant Profile Identifier** | **Factual GA at Trial Entry** | **17-OHPC** | **Placebo** | **17-OHPC** | **Placebo** | **GA at Entry (Weeks)** | **GA at Exit (Weeks)** | **Event[a]** | **GA at Entry (Weeks )** | **GA at Exit (Weeks )** | **Event[a]** |
| 1 | 16 | 1 | 1 | 1 | 1 | 16 | 19 | 1 | 16 | 19 | 1 |
| 2 | 20 | 1 | 1 | --- | --- | --- | --- | --- | --- | --- | --- |
| 3 | 16 | 0 | 1 | 0 | 1 | 16 | 36 | 1 | 16 | 37 | 0 |
| 4 | 20 | 0 | 1 | 0 | 1 | 20 | 36 | 1 | 20 | 37 | 0 |
| 5 | 16 | 0 | 1 | 0 | 1 | 16 | 35 | 1 | 16 | 37 | 0 |
| 6 | 20 | 0 | 1 | 0 | 1 | 20 | 35 | 1 | 20 | 37 | 0 |
| 7 | 16 | 0 | 0 | 0 | 0 | 16 | 37 | 0 | 16 | 37 | 0 |
| 8 | 20 | 0 | 0 | 0 | 0 | 20 | 37 | 0 | 20 | 37 | 0 |
| 9 | 16 | 0 | 0 | 0 | 0 | 16 | 37 | 0 | 16 | 37 | 0 |
| 10 | 20 | 0 | 0 | 0 | 0 | 20 | 37 | 0 | 20 | 37 | 0 |

GA = Gestational Age

[a] Event indicator values are 0, representing no composite outcome observed, and 1, representing composite outcome observed.
[b] The participant profile associated with participant profile identifier 2 experienced a miscarriage before 20 weeks' gestation, what would have been their factual gestational age at trial entry. So, while this profile is included for estimating the true counterfactual outcomes for randomization at 16 weeks' gestation, it is not included in the counterfactual outcomes for randomization at their factual gestational age at trial entry nor the analytic dataset for the extended Kaplan-Meier Estimator.

## Discussion

**Discussion S1.** Estimating the effect of 17-OHPC versus placebo in the presence of right-censored data without time-to-event estimators.

In the main text, we indicate that simple application of closed-cohort methods (e.g., proportions, logistic regression) should not be expected to return unbiased estimates of the total treatment effect in the presence of right censoring without additional analytic approaches or strong assumptions about censoring. We clarify those assumptions below, within the context of our hypothetical trial, and discuss additional analytic approaches that could be applied with closed-cohort methods to derive unbiased estimates of the total effect in the presence of right censoring.

While time-to-event methods are intuitive for complex treatment strategies, such as those where treatment cannot be distinguished at baseline, other approaches may be more straightforward for simpler cases, such as our motivating example, <u>if we are only interested in the cumulative incidence of the study outcome at the end of follow-up, **not the timing of events**</u>. In our hypothetical trial, treatment is defined at a single, aligned time point, and the event is defined at the end of pregnancy, meaning we are not necessarily interested in differences in the timing of the outcome. Thus, time-to-event methodology may not be necessary to estimate risks. We've intentionally chosen a simple example to help demonstrate the challenges inherent to using time-to-event estimators in pregnancy without competing complexities. However, as described below, closed-cohort approaches with inverse probability weighting could equivalently be used to estimate the intention-to-treat effect from our trial.

The directed acyclic graph (DAG) below depicts the key causal relationships in our hypothetical trial, where $A$ represents randomization to 17-OHPC versus placebo; $Y$ occurrence of delivery before 37 weeks' gestation, whether by miscarriage, stillbirth, or live birth; $C_1$ censoring due to LTFU at some time point after randomization; $L_0$ covariates measured at randomization; and $L_1$ the same covariates measured at some time point after randomization.

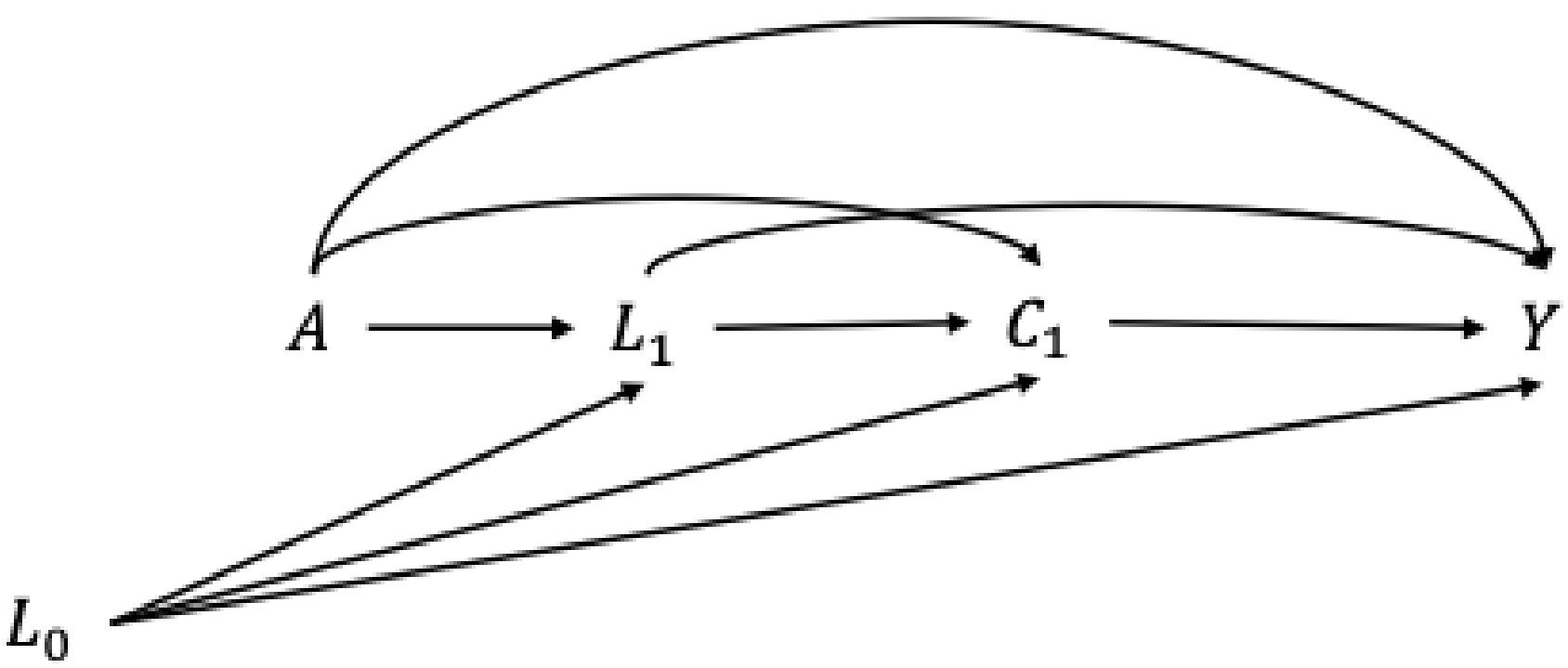

As stated in the main text, in the absence of censoring, we could estimate the total effect of $A$ on $Y$ without bias. In this setting, because we are interested in an end of follow-up outcome, rather than a survival outcome, individuals who reach term would be classified as not having the outcome (e.g., $Y = 0$), and we would not censor persons at term. We could then easily estimate the total effect of $A$ on $Y$ without bias. However, in the presence of loss to follow-up, we must carefully consider if we require additional approaches to account for selection from our baseline population to the population with observed outcomes. If we were willing to assume that censoring is completely uninformative, that is, unrelated to the risk of the outcome (i.e., $C \perp Y$),[1,2] the observed risk estimates among the subset of trial participants who were not LTFU would (in expectation) be unbiased estimates of the total effect.

However, as shown in the DAG above, it may be unreasonable to assume that LTFU is independent of the outcome. For example, we suppose that treatment affects the risk of LTFU, potentially by causing adverse events. Thus, conditioning on $C_1$ would result in collider stratification bias.[3–5] Further, we suppose that other risk factors affect both the risk of censoring and the outcome, potentially resulting in a differing distribution of effect measure modifiers between the original trial population and those who remain in the analytic sample (i.e., "type-2" selection bias).[1] In either scenario, we should not expect the observed risks among those who remain in our analytic sample to be an unbiased estimates of the total effect of treatment.

Inverse probability weights are a relatively straightforward approach to account for this potentially informative selection from the baseline trial population to our analytic sample. To derive unbiased risk estimates from the trial population that was not LTFU, we would need to remove all causal arrows from $C_1$ into $Y$. To do so, we would fit inverse probability weights using models for $C_1$ conditional on $A, L_0$, and $L_1$. In settings in which LTFU occurs at multiple times between baseline and end of follow-up, we could additionally fit time-varying inverse probability weights for censoring at all times between baseline and end of follow-up using pooled logistic regression models including a flexible function of time since baseline. Within the weighted pseudo-population, we could then fit outcome models conditional on randomization arm and any effect modifiers of interest, using linear regression, logistic regression, or other approaches depending on the nature of the outcome.

Of additional note, time-to-event estimators do not avoid issues related to informative censoring. Investigators must similarly account for such censoring through analytic approaches such as inverse probability of censoring weights.[2]

**Discussion S2.** “Healthy” live births as competing events.

The main text focuses on the importance of “reaching term” as a competing event in time-to-event analyses when using “time since study entry” as the analytic timescale. However, an additional competing event that rarely receives attention (but is applicable regardless of timescale) is “healthy” live birth, or, more accurately, live birth without the study outcome.[6]

To illustrate, consider that we are interested in estimating the effect of randomization to 17-OHPC versus placebo on the risk of preeclampsia at or before delivery. When applying time-to-event estimators like Kaplan-Meier, researchers frequently censor participants at the occurrence of a non-preeclamptic live birth, meaning that estimators of the risk of preeclampsia at delivery take the form $\prod_{k=0}^{K} \Pr(Y_{k+1} | Y_k = 0, C_k = 0, A = a)$ where $Y_k$represents preeclampsia by time $k$, $C_k$ represents birth by time $k$, and $A$ indicates randomization arm. If timing of non-preeclamptic live birth is random (meaning $Y_k^{a,c_{k-1}} \perp C_{k-1}$), or dependent only on randomization arm (meaning $Y_k^{a,c_{k-1}} \perp C_{k-1} \mid A$), then this quantity is equal to the counterfactual risk of preeclampsia by time k, under a joint intervention that sets randomization arm to $a$ and prevents non-preeclamptic delivery up to time k ($\mathrm{E}(Y_{k+1}^{a,\bar{c}_k})$). However, this controlled direct effect is likely not of interest, as we would not want to prevent non-preeclamptic deliveries. As discussed in the main text, this approach could be avoided by using approaches that allow for estimation of the total effect of randomization arm on outcomes up to time $k + 1$, such as the Aalen-Johansen estimator, or pooled logistic regression models for the discrete time hazard that set $Y = 0$ for all times after non-preeclamptic delivery.

To illustrate, consider two pregnancies enrolled at 9 weeks’ gestation: one experiences preeclampsia right before delivery at 42 weeks’ gestation, and the other results in live birth at 40 weeks’ without preeclampsia. If we censored the non-preeclamptic delivery at 40 weeks’ gestation, the time-to-event estimator would treat the person as remaining at risk of preeclampsia after delivery, and remaining observations would be upweighted to reflect the censored non-preeclamptic delivery (shown below).

Formulated this way, a “healthy” live birth may be considered an “absorbing state”, or an outcome that someone cannot transition out of.[7–11] Specifically, once someone delivers without preeclampsia, they are no longer at risk of developing pre-delivery preeclampsia (or any other delivery-related outcome). If we were to treat the end of that person’s follow-up as a censoring event, the analysis would imply the opposite: that the pregnancy is still at risk.

A person who experiences a non-preeclamptic live birth 40 weeks of gestation cannot experience pre-delivery preeclampsia at 42 weeks of gestation.

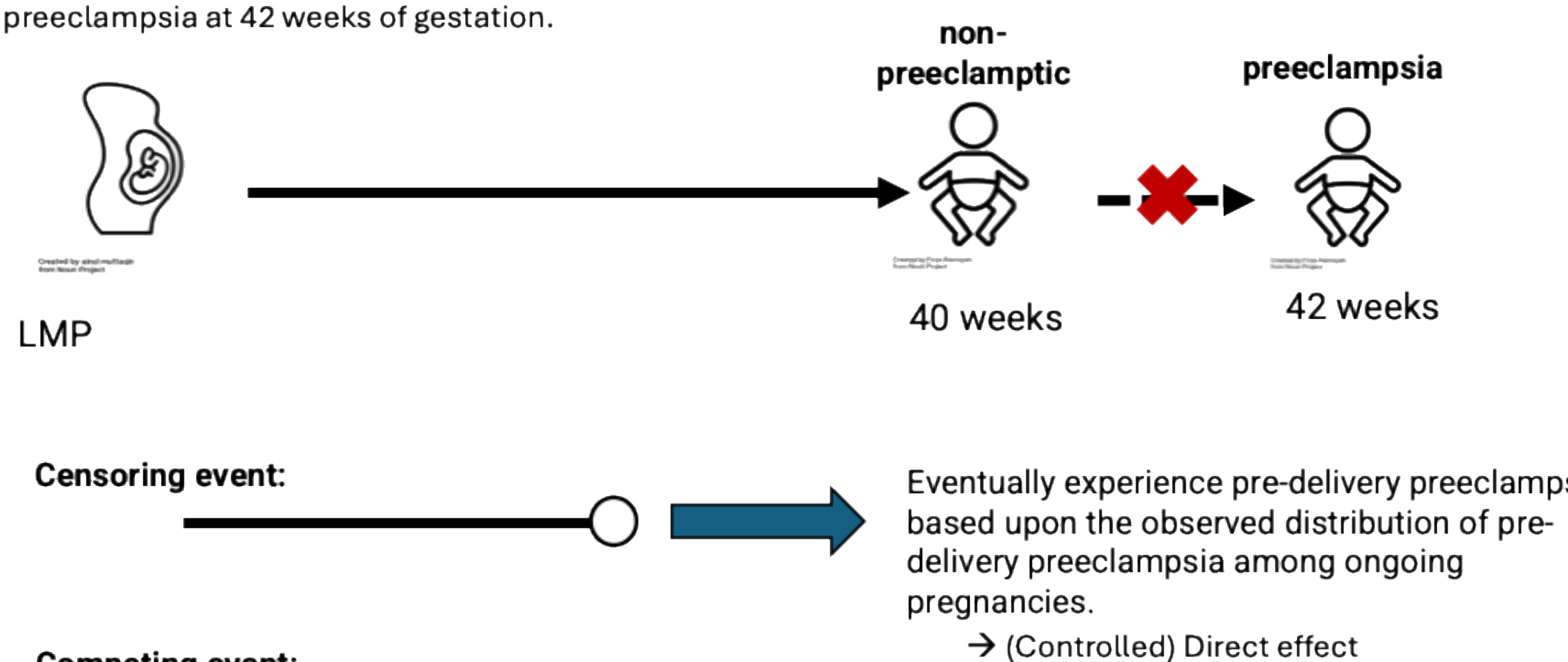

**Discussion S3.** Additional detail on bias due to variable left truncation.

As discussed in the main text, if investigators allow participants to enter their study within a range of gestational ages but conduct time-to-event analyses on the "gestational age" timescale, their study sample is subject to variable left truncation with regard to their target trial. That is, the start of follow-up varies across participants according to their factual gestational age at study entry.[12]

Bias due to variable left truncation can be illustrated with directed acyclic graphs (DAGs).[13] **Figure (A)** illustrates a simplified DAG for our hypothetical trial if we use the "time since study entry" timescale. In this DAG, there are no open backdoor paths between treatment (randomization at 16-20 weeks' gestation) and the outcome, so there is no bias in our intention-to-treat estimates of the total effect. However, using gestational age as the analytic timescale for an intention-to-treat analysis targets the effect of randomization at 16 weeks' gestation on our outcome. As shown in **Figure (B)**, trial entry at 17 weeks' gestation is a collider, affected by both remaining pregnant and prior randomization (since randomization at earlier weeks precludes randomization at later weeks).[1,4,5] The same bias could occur if gestational age at trial entry is affected by other causes of the outcome.[2,14–17] For example, persons with lower healthcare engagement may enroll later in pregnancy, which may also be related to their risk of the outcome. Estimating the effect of 17-OHPC initiation at 16 weeks' gestation on our composite outcome requires closing open paths between late entry (selection) and the outcome.

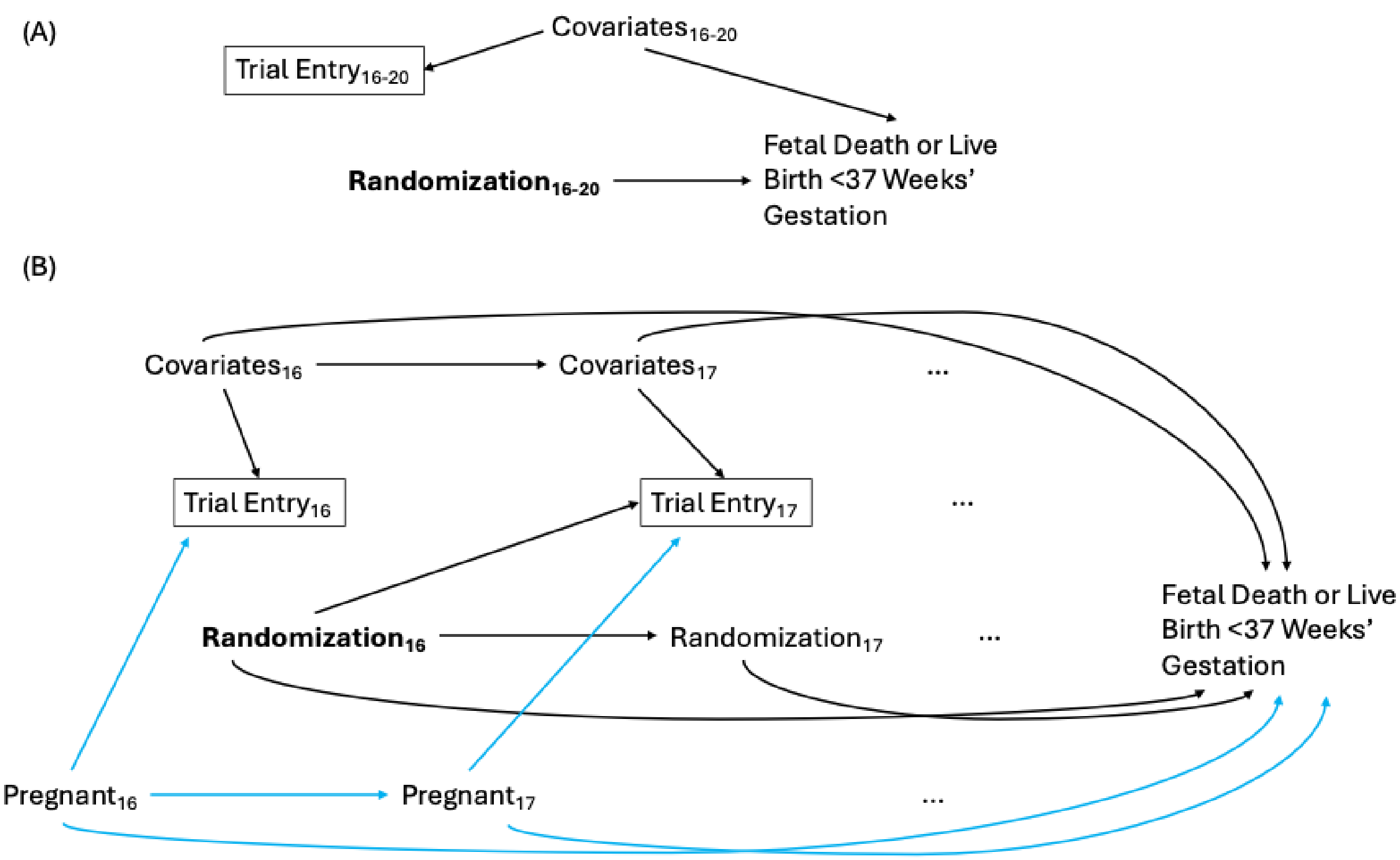

**Discussion S4.** Overview of methodologies to account for informative late entry.

As discussed in the main text, if investigators allow participants to enter their study within a range of gestational ages but conduct time-to-event analyses on the "gestational age" timescale, their study targets the treatment effect if, counter to fact, all participants who factually enrolled in the trial had actually entered the study at a specific gestational age, for example 16 weeks' gestation. In implementing such an analysis, it can be said that the study sample is subject to variable left truncation with regard to the target trial. If this left truncation is informative (i.e., related to the risk of the outcome either directly or through other variables such as healthcare engagement), investigators must implement methodology that can account for this potential source of bias.[18,19]

An example of one such approach is *g*-computation. Briefly, we could use the observed data from persons enrolled at gestational week 16 to build predictive models for their counterfactual outcomes conditional on measured variables. We could then predict the counterfactual risks for the entire study population based on their covariate distribution if, counter to fact, all had enrolled at the specified gestational age. However, implementing this analysis requires strong, untestable assumptions. First, we must assume no measurement error: the covariates measured at someone's factual gestational age at study entry perfectly represent those that would have been observed had they, potentially counter to fact, enrolled at 16 weeks' gestation. This may be violated if covariate values in the window between 16 weeks and later enrollment are themselves affected by exposure. Furthermore, we must be willing to assume positivity (i.e., everyone had a non-zero probability of enrolling at that gestational age), exchangeability (i.e., no unmeasured baseline or time-varying confounding, which may be unreasonable if you cannot measure all predictors of remaining pregnant [required for trial entry]), and no model misspecification.[20] Finally, the target population is not well-defined in this analysis:[21] the population of pregnancies that would have been observed had follow-up uniformly started at 16 weeks' gestation is unknown.[1]

Other approaches to informative left truncation exist beyond the outcome-modeling strategy above, though an exhaustive review is outside the scope of this manuscript.[14–17] We caution investigators that additional assumptions may be necessary depending on the approach applied. For example, inverse probability weights can be used to adjust for informative late entry.[22–24] However, including a person's observed risks after treatment at 20 weeks' gestation in the outcome model may imply a higher adherence to 17-OHPC than would have been observed had everyone, counter to fact, enrolled at 16 weeks' gestation. Further, many of these methods (e.g., "extended" Kaplan-Meier) assume a constant hazard or event rate after treatment initiation, which may be unreasonable.[25] Finally, investigators must carefully consider how to handle sparse data at earlier gestational ages at study entry.[26,27]